\documentstyle[preprint,prc,aps,floats,epsf]{revtex}

\begin{document}
%
%
\def\symdef#1#2{\def#1{#2}}
\symdef{\alphabar}{\overline\alpha}
\symdef{\alphabarp}{\overline\alpha\,{}'}
\symdef{\Azero}{A_0}

\symdef{\bzero}{b_0}

\symdef{\Cs}{C_{\rm s}}
\symdef{\cs}{c_{\rm s}}
\symdef{\Cv}{C_{\rm v}}
\symdef{\cv}{c_{\rm v}}

\symdef\dalem{\frame{\phantom{\rule{8pt}{8pt}}}}
\symdef{\Deltaevac}{\Delta{\cal E}_{\rm vac}}

\symdef{\ed}{{\cal E}}
\symdef{\edk}{{\cal E}_k}
\symdef{\edkzero}{{\cal E}_{k0}}
\symdef{\edv}{{\cal E}_{\rm v}}
\symdef{\edvphi}{{\cal E}_{{\rm v}\Phi}}
\symdef{\edvphizero}{{\cal E}_{{\rm v}\Phi 0}}
\symdef{\edzero}{{\cal E}_{0}}
\symdef{\Efermistar}{E_{{\scriptscriptstyle \rm F}}^\ast}
\symdef{\Efermistarzero}{E_{{\scriptscriptstyle \rm F}0}^\ast}
\symdef{\etabar}{\overline\eta}
\symdef{\ezero}{e_0}

\symdef{\fomega}{f_\omegav}
\symdef{\fpi}{f_\pi}

\symdef{\gA}{g_A}
\symdef{\gomega}{g_\omegav}
\symdef{\gpi}{g_\pi}
\symdef{\grho}{g_\rho}
\symdef{\gs}{g_{\rm s}}
\symdef{\gv}{g_{\rm v}}

\symdef{\fm}{\mbox{\,fm}}

\symdef{\infm}{\mbox{\,fm$^{-1}$}}

\symdef{\kappabar}{\overline\kappa}
\symdef{\kfermi}{k_{{\scriptscriptstyle \rm F}}}
\symdef{\kfermizero}{k_{{\scriptscriptstyle \rm F}0}}
\symdef{\Kzero}{K_0}

\symdef{\lambdabar}{\overline\lambda}
\symdef{\lzero}{l_{0}}

\symdef{\MeV}{\mbox{\,MeV}}
\symdef{\momega}{m_\omegav}
\symdef{\mpi}{m_\pi}
\symdef{\mrho}{m_\rho}
\symdef{\ms}{m_{\rm s}}
\symdef{\Mstar}{M^\ast}
\symdef{\Mstarzero}{M^\ast_0}
\symdef{\mv}{m_{\rm v}}
\symdef{\mzero}{{\rm v}_{0}}

\symdef{\Nbar}{\overline N}

\symdef{\omegaV}{V}
\symdef{\omegav}{{\rm v}}

\symdef{\Phizero}{\Phi_0}
\symdef{\psibar}{\overline\psi}
\symdef{\psidagger}{\psi^\dagger}
\symdef{\pvec}{{\bf p}}

\symdef{\rhoB}{\rho_{{\scriptscriptstyle \rm B}}}
\symdef{\rhoBzero}{\rho_{{\scriptscriptstyle \rm B}0}}
\symdef{\rhos}{\rho_{{\scriptstyle \rm s}}}
\symdef{\rhospzero}{\rho'_{{\scriptstyle {\rm s} 0}}}
\symdef{\rhoszero}{\rho_{{\scriptstyle {\rm s}0}}}
\symdef{\rhozero}{\rho_0}

\symdef{\Szero}{S_0}

\symdef{\Tr}{{\rm Tr\,}}

\symdef{\umu}{u^\mu}
\symdef{\Ualpha}{U_{\alpha}}
\symdef{\Ueff}{U_{\rm eff}}
\symdef{\Uzero}{U_0}
\symdef{\Uzerop}{U_0'}
\symdef{\Uzeropp}{U_0''}

\symdef{\vecalpha}{{\bbox{\alpha}}}
\symdef{\veccdot}{{\bbox{\cdot}}}
\symdef{\vecnabla}{{\bbox{\nabla}}}
\symdef{\vecpi}{{\bbox{\pi}}}
\symdef{\vectau}{{\bbox{\tau}}}
\symdef{\vecx}{{\bf x}}
\symdef{\Vzero}{V_0}

\symdef{\wzero}{w_0}
\symdef{\Wzero}{W_0}

\symdef{\zetabar}{\overline\zeta}
%
%
%
%

\draft

\preprint{\vbox{\hfill IU/NTC\ \ 96--12\\ \null\hfill NUC--MINN-96/19-T}}

\title{Vacuum Nucleon Loops and Naturalness}

\author{R. J. Furnstahl}
\address{Department of Physics \\
         The Ohio State University,\ \ Columbus, OH\ \ 43210}
\author{Brian D. Serot}
\address{Department of Physics and Nuclear Theory Center \\
         Indiana University,\ \ Bloomington, IN\ \ 47405}
\author{Hua-Bin Tang}
\address{Department of Physics \\
         University of Minnesota,\ \ Minneapolis, MN\ \ 55455}
%

%
\date{October, 1996}
\maketitle
\begin{abstract}
Phenomenological studies support the applicability of naturalness
and naive dimensional analysis to hadronic effective lagrangians
for nuclei.
However, one-baryon-loop vacuum contributions in renormalizable
models give rise to unnatural coefficients, 
which indicates that the quantum vacuum
is not described adequately.
The effective lagrangian framework
accommodates a more general characterization of vacuum contributions
without reference to a Dirac sea of nucleons.
\end{abstract}


Variations of the Walecka model are widely applied
in descriptions of nuclear structure and 
reactions \cite{SEROT86,REINHARD89,GAMBHIR90,SEROT92}.
These models are appealing both for their phenomenological success
and for the simplicity and economy of the physics at the mean-field level.
Most practical applications used for quantitative nuclear structure
phenomenology include contributions from valence
nucleons only; the framework is often labeled a ``no sea'' 
approximation to indicate that the Dirac sea of nucleons is 
neglected \cite{REINHARD89}. 

The approach to the relativistic nuclear many-body problem known as quantum
hadrodynamics (QHD) was originally based on renormalizable field 
theory \cite{SEROT86,SEROT92}, so that systematic calculations
with a finite number of parameters are possible, at least in principle.
In particular, one-baryon-loop vacuum effects can be included as a way to
add vacuum dynamics to the ``no sea'' physics.
These effects have a natural interpretation as the response 
of a filled Dirac sea of nucleons to the presence of valence 
nucleons\cite{SEROT86}.
The resulting ``relativistic Hartree approximation'' (RHA), however,
does not provide an acceptable description of the properties of finite
nuclei, at least by the standards of modern successful mean-field 
models \cite{FOX89,FURNSTAHL89,FURNSTAHL90,FURNSTAHL93,FURNSTAHL95}.
  
The deficiencies of the RHA could imply that one simply needs
to work harder to describe vacuum dynamics within the QHD framework.
It is known, however, that including higher-order corrections in a simple
loop expansion makes the phenomenology worse \cite{FURNSTAHL89}, so that
any improvement (if it exists) would require significantly more complicated
diagrams, such as those involving vertex corrections or short-range 
correlations \cite{ALLENDES92,SEROT95}.
Moreover, one would like to structure the hadronic many-body framework so 
that the appealing, intuitive, and successful mean-field theory is indeed
a good starting point.
This suggests an effective field theory (EFT) approach based on the 
``modern'' viewpoint of nonrenormalizable 
lagrangians \cite{WEINBERG79,LEPAGE89,POLCHINSKI92,GEORGI93,BALL94,WEINBERG95}.
This viewpoint is well known in the particle physics community (for systems
at zero density) and is adopted in applications
of chiral perturbation theory to meson--meson, meson--nucleon, and
nucleon--nucleon scattering.
However, it is not well known or applied in the community of nuclear
physicists who use relativistic models for nuclear structure.

From the EFT perspective, a description including only valence nucleons
and classical (``mean'') meson fields can {\it still\/} incorporate vacuum 
dynamics, hadron compositeness, and many-body correlations, albeit 
approximately \cite{FURNSTAHL95,FURNSTAHL96a,FURNSTAHL96b}.
The price to be paid is that all possible terms consistent with the
underlying symmetries (excluding redundancies) must appear in the effective
lagrangian.
Nevertheless, by relying on the concept of ``naturalness'' (as defined
below), it is possible to systematically truncate the effective lagrangian,
leaving only a finite number of unknown parameters;
moreover, recent fits to empirical nuclear properties using this framework
give strong evidence that the model parameters are indeed 
natural \cite{FURNSTAHL95,FRIAR96,FURNSTAHL96a,FURNSTAHL96b}.

From this point of view, the RHA in renormalizable QHD models
is simply one specific prescription for determining
an infinite number of parameters in the effective theory, namely, the 
coefficients in the scalar effective potential
and of terms involving derivatives of the boson fields.
Here we assess the relevance of the RHA prescription by examining the
size of these coefficients.
We find that the RHA leads to unnaturally large coefficients, in disagreement
with results obtained from fits to empirical nuclear 
properties \cite{FURNSTAHL96b}. 
This implies that although it may be possible to {\it explicitly\/} include
vacuum dynamics by calculating baryon vacuum loops, it is much more
efficient to include them {\it implicitly\/} in the small number of
natural parameters contained in the truncated effective lagrangian.

There have been more formal criticisms of the RHA:
(1) the RHA vacuum contributions violate $N_c$ counting rules motivated from
quantum chromodynamics (QCD) \cite{COHEN89,KIRITSIS89},
(2) the RHA neglects the compositeness of the 
nucleons \cite{BRODSKY84,NEGELE86,COHEN90}, and
(3) the treatment of the vacuum in terms of $N\Nbar$ pairs alone is
simply wrong, or at best incomplete \cite{BRODSKY84,COHEN90,%
BLESZYNSKI87,BANERJEE89,JAROSZEWICZ91}.
It seems much more compelling to us that one should avoid explicit
baryon-loop calculations simply because the {\it empirical properties
of nuclei\/} show that dynamical vacuum effects are quite modest and can be
described with a few adjustable parameters.

Georgi and Manohar\cite{GEORGI84b,GEORGI93} have proposed
a naive dimensional analysis (NDA) for assigning a coefficient of the 
appropriate size to any term in an effective lagrangian for the strong
interaction.
This NDA has been extended to effective hadronic lagrangians for nuclei
\cite{FRIAR96,FURNSTAHL96b}. 
The basic assumption of ``naturalness'' is that once the appropriate
dimensional scales have been extracted using NDA, 
the remaining overall dimensionless
coefficients should all be of order unity.
For the strong interaction, there are two relevant scales:
the pion-decay constant $\fpi \approx 93~\mbox{MeV}$ 
and a larger scale $0.5 \alt \Lambda \alt 1~\mbox{GeV}$, 
which characterizes the mass scale of physics beyond Goldstone bosons.
The NDA rules prescribe how these scales should appear in a given 
term in the lagrangian density:
\begin{description}
\item[1.] Include a factor of $1/\fpi$ for each strongly interacting
        field.
\item[2.] Assign an overall normalization factor of $\fpi^2 \Lambda^2$.
\item[3.] Multiply by factors of $1/\Lambda$ to achieve dimension (mass)$^4$.
\item[4.] Include appropriate counting factors (such as $1/n!$ for $\phi^n$). 
\end{description}
The appropriate mass for $\Lambda$
might be the nucleon mass $M$ or a non-Goldstone
boson mass; the difference is not important for 
numerical assessments of naturalness,
but will be relevant for the $N_c$ counting arguments considered later.

As an example of the NDA prescription,
a term in the scalar effective potential takes the form
\begin{equation}
  \kappa_n {1 \over n!} \fpi^2 \Lambda^2 
    \biggl(  {\phi\over \fpi} \biggr)^n
    \ . \label{eq:generic}  
\end{equation}
The coupling constant $\kappa_n$ is dimensionless 
and of $O(1)$ if naturalness holds.
Until one can derive the effective lagrangian from QCD, the naturalness 
assumption must be checked by fitting to empirical nuclear data.
Such fits give strong support for naturalness \cite{FRIAR96,FURNSTAHL96b}.

We can assess the naturalness of the RHA vacuum contributions by matching
the results in a renormalizable model
to an effective mean-field theory in which the Dirac sea
degrees of freedom are excluded by construction.
The matching to the RHA is conveniently made with an effective
action, in which the Dirac sea contribution is easily isolated.
(See, for example, Ref.~\cite{FURNSTAHL95}.)
The one-loop Dirac sea effective action $\Gamma'$ in models
with Yukawa couplings to the nucleon takes the (unrenormalized) form
\begin{equation}
  \Gamma' \equiv -i\hbar\, \Tr\ln (i\gamma^\mu\partial_\mu
      - \Mstar - \gv \gamma^\mu V_\mu) \ ,  \label{eq:trln}
\end{equation}
where $\Mstar \equiv M - \gs\phi$, $\phi$ and $V^\mu$ are neutral
scalar and vector fields, $\gs$ and $\gv$ are their couplings to
the nucleon, 
and the trace is over spatial and internal variables.
For time-independent background meson fields, the effective action
is proportional to the energy.%
\footnote{The contribution to the energy from
Eq.~(\ref{eq:trln}) can be written more transparently as a sum over 
single-particle energies of
occupied states in the filled Dirac sea \cite{SEROT86}.  
In nuclear matter, this (unrenormalized) energy density is
$\Delta{\cal E} = V^{-1}\sum_{{\bf k},\lambda}
   \bigl[({\bf k}^2 + \Mstar{}^2)^{1/2} - 
        ({\bf k}^2 + M^2)^{1/2}\bigr] $.} 

The expression in Eq.~(\ref{eq:trln}) can be 
renormalized and evaluated as a derivative
expansion \cite{AITCHISON84} in the $\phi$ and $V^\mu$ fields:
\begin{eqnarray}
  \Gamma'_{\rm ren} &=& \int\! d^4x\ [
   - \Ueff(\phi) + {1\over 2}Z_{\rm 1s}(\phi) \partial_\mu\phi \partial^\mu\phi
     + {1\over 2}Z_{\rm 2s}(\phi)(\dalem\,\phi)^2   \nonumber \\
     & & \null + {1\over 4}Z_{\rm 1v}(\phi) F_{\mu\nu}F^{\mu\nu}
       + {1\over 2}Z_{\rm 2v}(\phi)(\partial_\alpha F^{\alpha\mu})
       (\partial^\beta F_{\beta\mu}) + O(1/ \Mstar{}^{3})]
       \ .  \label{eq:derivexp} 
\end{eqnarray} 
This expansion in inverse powers of $\Mstar$ 
converges rapidly for finite nuclei \cite{PERRY86,WASSON88,CARO96,FERREE93}.
The effective potential $\Ueff$ and the
coefficient functions $Z_i$ can be further expanded in (infinite) polynomials
in $\phi$ with well-behaved coefficients;  
that is, there is a local expansion of $\Gamma'$, which can be absorbed
into an effective lagrangian for nuclei.
Thus the finite Dirac sea contribution can be reproduced by an 
effective lagrangian treated at the mean-field level with only valence 
nucleons, as long as all possible terms (generally nonrenormalizable) are
included.

The effective potential $\Ueff(\phi)$ is found by evaluating
the trace in Eq.~(\ref{eq:trln}) 
with {\it constant\/} fields.  This expression is divergent
and must be regularized and renormalized to obtain a finite result, which 
takes the general form \cite{SEROT86}
\begin{equation}
  \Ueff(\phi) = -{\gamma\over (4\pi)^2}
      \Bigl( \Mstar{}^4\ln {\Mstar\over\mu} + \sum_{n=0}^4
             \alpha_n M^{4-n}(\gs\phi)^n
       \Bigr)  \ ,  \label{eq:ueff}
\end{equation}
where $\gamma$ is the spin-isospin degeneracy and the $\alpha_n$
are dimensionless constants.  
The scale parameter $\mu$ is typically
chosen to be $M$.
The counterterms $\alpha_0$ and
$\alpha_1$ are fixed by requiring $U_{\rm eff}$ to be zero and
a minimum in the vacuum ($\phi=0$).  The others are fixed by
prescription.  
Note that in a renormalizable model, only the first four powers of
$\phi$ are available as counterterms.  In an effective lagrangian, however,
{\it all\/} powers are present.

The most common prescription has been to choose the $\alpha_n$ to cancel
the first four powers of $\phi$ appearing in the expansion of the
logarithm \cite{SEROT86}.
One finds in this case (for $\gamma=4$)
 \begin{eqnarray}
   \Ueff(\phi)      &=&
      -{1\over 4\pi^2}\Big[
      (M-\Phi)^4 \ln(1-{\Phi\over M}) + M^3 \Phi 
      - {7\over 2}M^2 \Phi^2 + {13\over 3}M\Phi^3
      - {25\over 12}\Phi^4 \Big] \nonumber \\[4pt]
      &=&
      {M^4\over 4\pi^2}\Bigl\{
      {\Phi^5\over 5 M^5} + {\Phi^6\over 30M^6}
      + {\Phi^7 \over 105 M^7} + \cdots
      + {4!(n-5)!\over n!} {\Phi^n\over M^n} + \cdots
      \Bigr\}  \ ,  \label{eq:deltaE}
 \end{eqnarray}
where $\Phi \equiv \gs\phi$.
When $\phi$ (or $\Mstar$) is determined at each density by minimization, 
$\Ueff(\phi)$ is the finite shift in the baryon zero-point energy
that occurs at finite density and is analogous to the ``Casimir energy''
that arises in quantum electrodynamics.  

To evaluate the size of the one-loop vacuum correction, we apply the NDA.
Based on the scaling rules discussed above, a term of $O(\phi^5)$
should be scaled as
\begin{equation}
	{M^2\over 5! \fpi^3}\, \phi^5  \ , \label{eq:good}
\end{equation}
where we have associated $\Lambda$ with the nucleon mass $M$.
(See, however, the comments on $N_c$ scaling below.)
If this contribution is natural, any residual overall constant should
be of order unity.
However, if we perform a similar scaling on the leading term in
Eq.~(\ref{eq:deltaE}), we find
\begin{equation}
   {M^4\over 4 \pi^2}\, {\gs^5\phi^5\over 5 M^5} \longrightarrow
    {4\over 5}\,{M^2 \over \fpi^3}\, \phi^5
    = 96 \bigg( {M^2\over 5! \fpi^3}\, \phi^5 \bigg) \ ,
        \label{eq:bad}
\end{equation}
where we used $4\pi\fpi \approx M$ and $\gs \approx M/\fpi$.
Thus the one-baryon-loop contribution to the vacuum energy is
roughly two orders of magnitude {\em larger\/} than naturalness requires.
It is not hard to show from Eq.~(\ref{eq:deltaE}) that all higher
powers of $\Phi$ contain essentially the same large overall factor.

We can make
an instructive comparison of natural and unnatural coefficients 
using the linear sigma model, generalized to include a neutral
vector meson. 
Variations of this model have been used to investigate the role
of chiral symmetry
in nuclear structure \cite{MATSUI82,SEROT86,SEROT92}.%
\footnote{The limitations of this approach are discussed in 
Refs.~\cite{FURNSTAHL93} and \cite{FURNSTAHL95}.}
After following conventional procedures to introduce a nonzero
expectation value for the scalar field and then shifting the field,
we obtain the model lagrangian (with $\mpi=0$ for simplicity):
\begin{eqnarray}
  {\cal L}_{\sigma\omega} &=& \psibar \bigl[
    i \gamma_\mu \partial^\mu - \gv \gamma_\mu V^\mu
    - (M- \gpi \sigma) - i \gpi \gamma_5 \bbox{\tau\cdot\pi} \bigr] \psi 
      \nonumber \\
    \null & & \null + {1\over 2} (\partial_\mu \sigma \partial^\mu \sigma
        - m_\sigma^2 \sigma^2)
      - {1\over 4}(\partial_\mu V_\nu - \partial_\nu V_\mu)^2
      + {1\over 2}\mv^2 V_\mu V^\mu  \nonumber  \\
      & & \null + {1\over 2}\partial_\mu \bbox{\pi \cdot {}} \partial^\mu 
      \bbox{\pi}
      + \gpi {m_\sigma^2\over 2M} \sigma(\sigma^2 + \bbox{\pi}^2)
      - \gpi^2 {m_\sigma^2 \over 8M^2} (\sigma^2 + \bbox{\pi}^2)^2  \ ,
        \label{eq:sigom}
\end{eqnarray}
where the associations $\sigma \rightarrow \phi$,  $\gpi \rightarrow \gs$,
and $m_\sigma \rightarrow \ms$ should be made for our discussion.
Comparing to Eq.~(\ref{eq:generic}), with $\Lambda$ identified as $\ms$,
we find 
$\kappa_3 = -\kappa_4 = -3$, so that the nonlinear parameters are natural
at the mean-field level.%
\footnote{This is to be expected; the model builds in at tree level
a realization of QCD chiral symmetry and Goldstone-boson physics, which
is the same physics behind the NDA.}
However, if one includes the one-baryon-loop vacuum corrections,
renormalized in a fashion that preserves the chiral 
symmetry \cite{MATSUI82,FURNSTAHL93}, one finds {\it unnatural\/}
corrections to the cubic and quartic couplings: 
$\Delta\kappa_3 = 2 M^2 / \pi^2 \fpi^2 \approx 20$, 
$\Delta\kappa_4 = -8 M^2/ \pi^2 \fpi^2 \approx -80$.
The quintic and higher corrections are exactly the same as in 
Eq.~(\ref{eq:deltaE}).

These unnatural coefficients generate correspondingly large corrections 
to the conventional linear Walecka model 
mean-field theory (MFT).
If we adjust the model parameters 
to reproduce the standard nuclear matter properties in the RHA (equilibrium
at Fermi momentum $\kfermi^0 = 1.30\infm$ with a binding energy of 15.75 MeV),
the baryon effective mass at equilibrium becomes $\Mstar / M \approx 0.73$.
This translates into a change in the scalar potential $\Phi$ from
$430\MeV$ in the MFT to $250\MeV$ in the RHA.
This is a large effect, particularly since it is the $\phi^5$ term
that makes the difference.  When a general effective lagrangian is
fit to nuclear properties, terms of this order play essentially no
role \cite{FURNSTAHL96b}.

\begin{figure}[t]
 \setlength{\epsfxsize}{4.0in}
  \centerline{\epsffile{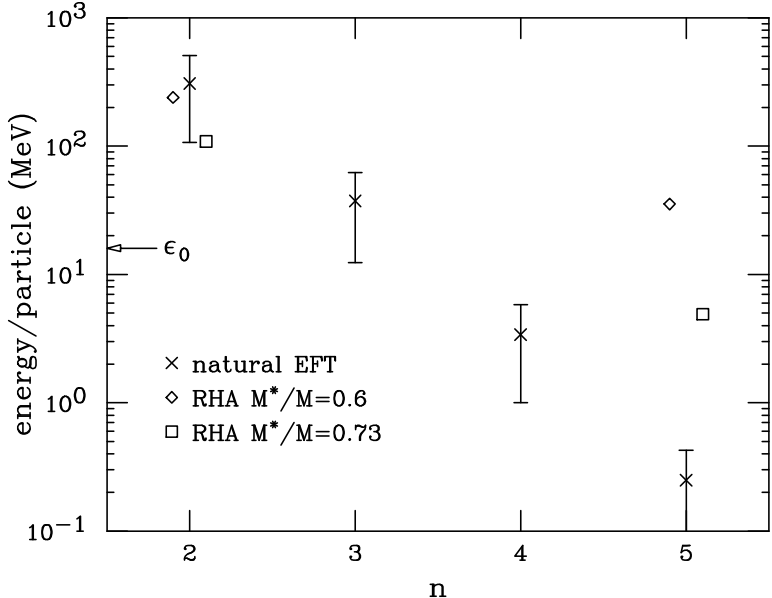}}
\vspace{.2in} \caption{Contributions to the scalar potential
 per particle in nuclear matter from the $n^{\rm th}$-order terms
 of the form $\Phi^n$ for the RHA model.  
The crosses are estimates based on Eq.~(\ref{eq:generic}).
The arrow indicates the total binding energy $\epsilon_0 =15.75~\mbox{MeV}$.}
 \label{fig:one}
\end{figure}

\begin{figure}[t]
 \setlength{\epsfxsize}{4.0in}
  \centerline{\epsffile{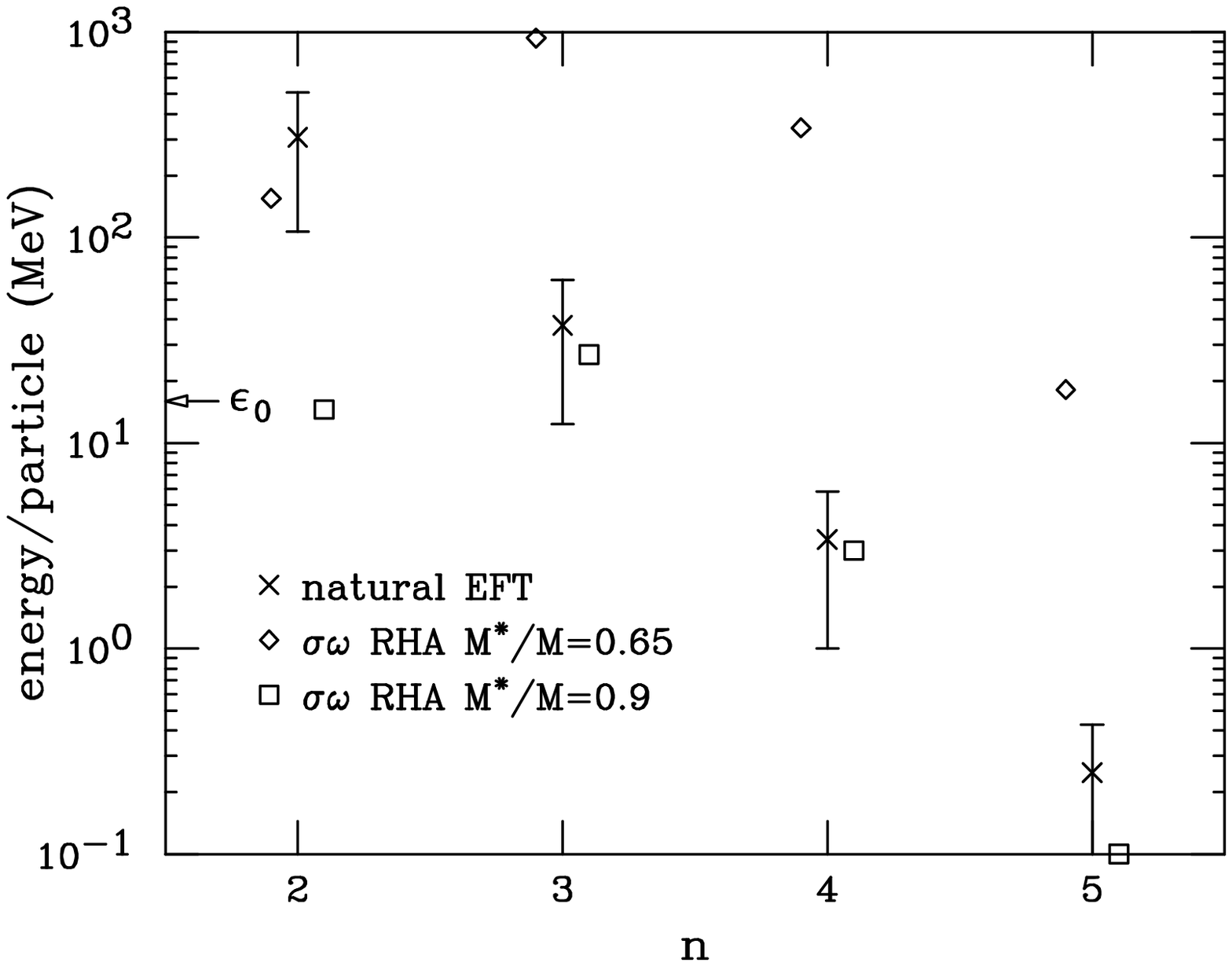}}
\vspace{.2in} \caption{Contributions to the scalar potential
 per particle in nuclear matter from the $n^{\rm th}$-order terms
 of the form $\Phi^n$ for the linear sigma model plus RHA.  
The crosses are estimates based on Eq.~(\ref{eq:generic}).
The arrow indicates the total binding energy $\epsilon_0 =15.75~\mbox{MeV}$.}
 \label{fig:two}
\end{figure}

Contributions to the scalar potential at equilibrium density
from $\Ueff(\phi)$ and from the
original tree-level lagrangians are plotted in 
Fig.~\ref{fig:one} for the Walecka model and in Fig.~\ref{fig:two}
for the linear sigma model.
In each figure, the crosses indicate contributions
to the energy density estimated using naturalness, with error bars 
allowing for a range of $\Lambda$ and $\Mstar/M$.
A steady decrease with increasing $n$ is evident, which motivates
the truncation of effective lagrangians with natural coefficients.
$\Mstar/M$ in natural models (and phenomenologically successful ones)
is typically between 0.60 and 0.66.
For those values, the RHA $O(\Phi^5)$ contribution would be as large as
a typical $O(\Phi^3)$ contribution in a natural model
(see Fig.~\ref{fig:one}) and would prevent a successful description of
nuclear structure.
In fact, the unnatural and unbalanced $O(\Phi^5)$ term drives $\Mstar/M$ 
to its self-consistent value of 0.73.
Higher-order contributions are essentially negligible, because of the
natural factor of $\phi/\fpi$ (and a combinatoric factor) that accompanies
each higher power.
In the linear sigma model, which has unnatural $O(\Phi^3)$
and $O(\Phi^4)$ contributions from the RHA, $\Mstar/M$ is driven to 0.9.

It is possible to devise a prescription \cite{RUDAZ92} that leads to 
an effective mass  $\Mstar/M$ and a nuclear compressibility
consistent with a reasonable (although not optimal) 
fit to properties of finite nuclei
(see Ref.~\cite{FURNSTAHL96a} for the criteria).
However, this requires choosing $\alpha_n$ coefficients in 
Eq.~(\ref{eq:ueff})
to achieve sensitive cancellations between terms of different order
in the effective potential,
so as to neutralize the effect of the unnatural $\phi^5$ contribution.
The unnaturalness of the vacuum nucleon loop contributions 
is generally characterized by the changes in the coefficients
that accompany $O(1)$ changes in the 
scale $\mu$ [say, from $\mu$ to $e\mu$ in Eq.~(\ref{eq:ueff})],
as in Ref.~\cite{GEORGI84b}.
The subsequent changes in the $\alpha_n$ coefficients are obtained by
expanding
$\gamma\Mstar{}^4/(4\pi^2) $.
These changes are large compared to the natural size implied by
Eq.~(\ref{eq:generic}).

Based on the strong empirical evidence for naturalness, we conclude that the
treatment of the quantum vacuum at the one-baryon-loop level is, 
at best, inadequate.
Although the concept of a Dirac sea is compelling for nuclear physicists
because of the analogy to the Fermi sea, the explicit calculation of these
effects prejudices the description of the vacuum dynamics and (to date)
has not yielded results consistent with nuclear structure 
phenomenology.
Moreover, the self-consistent, valence-nucleon-only theory is covariant,
causal, and internally consistent by itself, and the empirical evidence
shows that the vacuum degrees of freedom can be included implicitly by a
small number of local interactions among the mesons and valence nucleons.
Note that the omission of explicit dynamical contributions from the Dirac
sea does {\it not\/} mean that one can discard negative-energy solutions 
entirely; they 
must be retained to ensure the completeness of the Dirac wave functions
in certain calculations of density-dependent effects (for example, linear
response) \cite{FURNSTAHL87,DAWSON90}.

It should also be emphasized that an effective lagrangian allows for a more 
general characterization of the vacuum dynamics than that arising from
baryon loops.
The explicit calculation of counterterms in the effective lagrangian
is unnecessary,
since the end result is simply an infinite polynomial in
the scalar field, with finite, unknown, and apparently natural coefficients
arising from the underlying dynamics of QCD.
To have predictive power, one must rely either on the truncation scheme
provided by naturalness (see Ref.~\cite{FURNSTAHL96b}), 
so that only a small, finite number of unknown coefficients
are relevant, or on some other dynamics to constrain the form of the
renormalized scalar potential.
For example, a simple model is used in \cite{FURNSTAHL95} to show how the
broken scale invariance of QCD leads to dynamical constraints on the scalar 
potential, and fits to the properties of finite nuclei also generate
coefficients that are natural.
  
Another potential difficulty in the explicit calculation of baryon vacuum
loops is that RHA vacuum contributions violate
$N_c$ counting rules motivated by consistency with
QCD \cite{COHEN89,KIRITSIS89}.
The expected $N_c$ scaling of the
coupling for an $n$-meson vertex is $O(N_c^{1-n/2})$ \cite{WITTEN79}.
The $N_c$ scaling property of a vertex in an effective lagrangian
or the RHA
is established with the associations $M \propto O(N_c)$, $\ms \propto O(1)$,
and $\gs^2 \propto O(N_c)$ \cite{WITTEN79}.  
[Note also that $\fpi \propto O(N_c^{1/2})$.]
Thus one may simply inspect the coefficient of $\phi^n$.  
For the RHA, the result is $O(N_c^{4-n/2})$,
which exceeds the expected scaling by a factor of
$N_c^3$ \cite{COHEN89,KIRITSIS89}. 

In contrast, the effective lagrangian approach, with our definition of naive 
dimensional analysis and naturalness, is consistent
with the scaling expected from QCD if one associates
$\Lambda$ in the $\phi^n$ terms with $\ms$ (and not $M$).    
For example, applying the $N_c$ scaling rules to the coefficient of 
$\phi^5$ produces
\begin{equation}
  \kappa_5 {1\over 5!} \fpi^2\Lambda^2 {\phi^5\over \fpi^5}
  \longrightarrow \kappa_5 {1\over 5!} (\ms^2\phi^2) {\phi^3\over \fpi^3}
    \longrightarrow  O(1) \times O(N_c^{-3/2})
        \propto O(N_c^{-3/2}) \ ,
\end{equation}
which is consistent with the usual $N_c$ counting.
It is important here that the normalization factor $\Lambda^2\fpi^2$
comes from the meson mass term,
with the meson mass $\ms$ being $O(1)$ rather than $O(N_c)$.
Thus $N_c$ counting implies that the vacuum response is $\overline qq$
in nature; the response of a $N\Nbar$ vacuum is unlikely
to agree with the large $N_c$ limit.

Finally, we comment on the statement often made      
or implied in the literature that hadronic
field theories treat nucleons as point particles and so
do not account for the compositeness of nucleons.
The use of effective lagrangians with fields for composite particles 
is by now well established (e.g., chiral perturbation theory
with nucleons and pions).
In Ref.~\cite{FURNSTAHL96b}, the electromagnetic structure of the nucleon
manifestly appears at the mean-field level using a hadronic effective
lagrangian for nuclei.
The key features are the inclusion of
nonrenormalizable interaction terms and the use of a derivative expansion
to incorporate the nucleon compositeness; indeed, the allowed freedom in
the definition of the boson fields shows that even renormalizable terms
like $\phi^3$ and $\phi^4$ implicitly include the effects of nucleon
substructure.
Thus, the deficiencies of the RHA are not intrinsic
to the use of nucleons (and the Dirac equation) to describe nuclei,
but arise because the implied vacuum dynamics is incorrect or incomplete.    

In summary, we have examined the relativistic Hartree approximation (RHA) to 
renormalizable hadronic field theories in the context of modern effective
field theory.
The parameters obtained in phenomenologically successful mean-field models
of finite nuclei exhibit naturalness, as anticipated from naive dimensional
analysis of the strong interaction.
However, the parameters implied from vacuum loops in the RHA are 
{\it unnatural}.
Since the phenomenological parameters in successful models implicitly include
the effects of vacuum dynamics, we conclude that the explicit treatment of
the vacuum in the RHA, which involves the Dirac sea of nucleons,
is inadequate.
In contrast, a natural effective model with only valence nucleons, 
including all possible (nonredundant) terms, is consistent with nucleon
compositeness, $N_c$ counting, and nuclear structure phenomenology.
Extensions to include higher-order vacuum loops will be discussed
in subsequent work.


\acknowledgments

We thank P. Ellis, S. Rudaz, and J. Rusnak for useful discussions.
This work was supported in part by the Department of Energy
under Contracts No.\ DE--FG02--87ER40365 and DE--FG02--87ER40328, the 
National Science Foundation
under Grants No.\ PHY--9511923 and PHY--9258270, and the A. P. 
Sloan Foundation.


\end{document}